\documentclass[epj]{svjour}
\usepackage{ulem}
\usepackage{amssymb}
\usepackage{graphicx}
\usepackage{epsfig}
\usepackage{subfigure}
\usepackage{verbatim}
\usepackage{color}
\usepackage{enumerate}
\usepackage{bm}

\begin{document}

\title{ {Formation} of Solitonic Bound State via Light-Matter Interaction}

\author{Priyam Das\inst{1,}\thanks{daspriyam3@gmail.com} \and Ayan Khan\inst{2,}\thanks{ayan.khan@bennett.edu.in} \and  Anirban Pathak\inst{3,}\thanks{anirban.pathak@jiit.ac.in}}
\institute{
   Department of Physics, Bankura Sammilani College, Kenduadihi, Bankura, WB-722101, India  \and
   Department of Physics, School of Engineering and Applied Sciences, Bennett University, Greater Noida, UP-201310, India \and
   Jaypee Institute of Information Technology, A-10, Sector-62, Noida, UP-201309, India
   }

\abstract{
Exchange of energy by means of light-matter interaction provides a new dimension to various nonlinear dynamical systems. Here, the effects of light-matter interaction are investigated for a situation, where two counter-propagating, orthogonally polarized laser pulses are incident on the atomic condensate. It's observed that a localized laser pulse profile can induce localized modes in Bose-Einstein condensate. A stability analysis performed using  Vakhitov-Kolokolov-like criterion has established that these localized modes are stable, when the atom-atom interaction is repulsive. The cooperative effects of light-matter interactions and atom-atom interactions on the Lieb-mode have been studied in the stable region through atomic dispersion, revealing the signature of bound state formation when the optical potential is P\"oschl-Teller type. The energy diagram also indicates a continuous transfer of energy from the laser pulses to the atoms as the light-matter interaction changes its sign.
}

\maketitle

\section{Introduction}
Matter-field interaction is in the heart of spectroscopy and quantum mechanics, Rayleigh scattering to Raman effect, Eddington's experimental verification of the general theory of relativity to the recent detection of gravitational waves in LIGO, all are manifestations of matter-field interaction. This is why  matter-field interaction has drawn considerable attention of the scientific community since long. In fact, it has also played a fundamental role in the development of quantum mechanics. Specifically, photoelectric effect and Compton effect are two nice examples of matter-filed interaction that played a pivotal role in the development of quantum physics. Later, the interest on matter-field interaction received a boost when laser was discovered in 1960s. Invention of laser helped us to develop new fields like nonlinear optics, quantum optics and atom optics, and thus to reveal true power and beauty of matter-field interaction through the appearance of several new phenomena (cf. \cite{withAKG} and references therein). In fact, the advent of laser also helped  in the experimental realization of Bose-Einstein condensates (BECs) as laser played a pivotal role in cooling and trapping \cite{MOT}. It also helped  in realizing ultra-cold gases and super solids and thus opened up a domain of recent  interest where one  studies interaction of light with ultra-cold matter \cite{ultracold1,ultracold2}.

This particular facet of the study of matter-field interaction is interesting for various reasons. For example,  in a BEC, it's impossible to distinguish the bosons and thus to identify- which boson (particle) has scattered a photon. Consequently, a collective scattering occurs and a correlation is developed between the bosons, which subsequently enhances the scattering and may lead to superradiance \cite{Superradiance1,Superradiance2,Superradiance3,Superradiance4,nagy2008self,das2016collectively} and other phenomena \cite{saidi2018ultrastrong,dimitrova,ma2017rabi}. In a set of recent works \cite{dimitrova,ritsch1,ritsch2}, the purview of BEC-light interaction has been extended beyond superradiance, and exciting new nonlinear dynamical phenomena have been reported. Particularly, in Ref. \cite{ritsch1}, a BEC was illuminated by two far off-resonant counter propagating non-interacting laser beams having orthogonal polarization, and it was theoretically observed that the spontaneous crystallization of light and ultra-cold atoms may happen. Specifically, in free space, periodic pattern formation for a BEC was reported in a novel regime which was not explored in the earlier works on the self-ordering effect \cite{selfordering1,selfordering2,selfordering3}. This work established that the possibilities of observing exciting new phenomena in the synthetic solid-state systems may be investigated through the quantum simulations with ultra-cold atoms in optical
lattices \cite{ritsch1}. Subsequently, the same system was numerically investigated in Ref. \cite{ritsch2} to reveal the growth dynamics involved in the self-ordering (spontaneous crystallization) process mentioned above. A similar system was also studied experimentally in \cite{dimitrova} by Dimitrova et al. Further, an analogous system was studied in Ref. \cite{schmidt2018} in the context of a one-dimensional Coulomb crystal.

The exchange of energy and momentum through coherent scattering of light from ultra-cold atoms are usually found to introduce different types of nonlinear dynamical effects. The richness of the subject has further been exposed through the investigation of opto-mechanical strain observed in the ultra-cold atomic cloud \cite{davidson}. Moreover, in the recent years, ultra-cold atomic gases have also been used as emulators for different condensed matter systems. The idea of emulation is also extended to understand the behavior of electrons in solids when subjected to a strong electric field by means of applying laser field on atomic condensate \cite{rajagopal1,rajagopal2}. These recent investigations \cite{dimitrova,ritsch1,ritsch2,schmidt2018}, motivated us to look into the interplay between matter-matter and light-matter interaction.

Specifically, we aim to investigate generation of localized structures in BEC under the influence of counter propagating non-interacting laser beams. The system is analogous to the physical system discussed in \cite{dimitrova,ritsch1,ritsch2}. However, we take a different approach and the region explored is  quite different and consequently, it is expected to reveal new insights. Specifically, in what follows, we investigate the effect of light-matter interaction, when laser pulses of localized nature are incident on the atomic condensate. We show that our theoretical analysis hints at the solitonic bound state formation for attractive light-matter interaction and repulsive atom-atom interaction. The validity of the obtained solution is checked by performing a stability analysis by using Vakhitov-Kolokolov-like (VK-like) criterion \cite{vakhitov1973stationary,kivshar1989,das2}.  The energy diagram also indicates a transfer of energy from photon to atom as the light-matter interaction turns repulsive from attractive.

The paper is organized as follows. In Section \ref{model}, we explicate the model where we briefly describe the dimensional reduction method to reduce the condensate from 3+1 to 1+1 dimension. Also, we illustrate the nature of the various types of the potentials present in the system. In Section \ref{solu}, we report the localized solution for the condensate and their region of stability. Furthermore, we probe the localized structure by examining its dispersion and matter-field energy exchange in Section \ref{ene}. Finally, the paper is concluded in  Section \ref{con}.


\section{The Model}\label{model}
We consider a trapped atomic BEC interacting with two counter propagating, orthogonally polarized laser beams. The EM field generates an optical potential for the atoms in the condensate, which in turn modifies the optical field. Consequently, the model system would allow light and matter to contentiously exchange energy and thus to dynamically modify each other.

The BEC can be treated within the purview of the mean-field formalism, by Gross-Pitaevskii (GP) equation \cite{gross,pitae} as
\begin{eqnarray}
\hskip-0.5cm
i\hbar\frac{\partial\Psi(\mathbf{r},t)}{\partial t} &=& \left[-\frac{\hbar^2}{2m}\nabla^2+U|\Psi(\mathbf{r},t)|^2 + V(\mathbf{r}) \right] \Psi(\mathbf{r},t) \nonumber \\ & &  - \mu \Psi(\mathbf{r},t) ,\label{3dgp}
\end{eqnarray}
where $U=4\pi\hbar^2a_s/m$, $a_s$ being the s-wave scattering length, $\mu$ is the chemical potential and $m$ being the mass of the atoms. Further, $V(\mathbf{r}) = V_{\mathrm{T}}(y,z) + V_{\mathrm{L}}(x)$ defines the external potentials, where, $V_{\mathrm{T}}(y,z)=\frac{1}{2} m \omega_{\perp}^2(y^2+z^2)$ refers to the transverse confinement with $\omega_{\perp}$ being the transverse trap frequency. Similarly, $V_{\mathrm{L}}(x)$ corresponds to the longitudinal confinement and described as $V_{\mathrm{L}}(x) = V_{\mathrm{HO}}(x) + V_{\mathrm{opt}}(x)$, where  $V_{\mathrm{HO}}(x)=\frac{1}{2} m \omega_{0}^2 x^2$ is a harmonic trap with $\omega_{0}$ being the longitudinal trap frequency, and $V_{\mathrm{opt}}$ being the optical potential created due to the presence of the counter-propagating laser beams. It is worth highlighting that in order to get a quasi-one dimensional scenario, in what follows, we would consider that the transverse trapping frequency is much stronger compared to the longitudinal frequency ($\omega_{\perp} >> \omega_{0}$). This implies that the interaction energy of the atoms is much less than the kinetic energy in the transverse direction.  Consequently,  it is possible to reduce Eq.(\ref{3dgp}) from $3+1$ dimension to $1+1$ dimension assuming that the chemical potential of the condensate is much smaller than the transverse trap frequency, $\mu<<\hbar\omega_{\perp}$. In order to reduce Eq.(\ref{3dgp}) to  the corresponding quasi one-dimensional case, we have made use of the following ansatz, too
\begin{eqnarray}
\Psi(\mathbf{r}, t) &=& \frac{1}{\sqrt{2\pi a_B}a_{\perp}}\psi\left(\frac{x}{a_{\perp}},\omega_{\perp}t\right) e^{\left(-i\omega_{\perp}t-\frac{y^2+z^2}{2a_{\perp}^2}\right)},\label{ansatz}
\end{eqnarray}
where $a_B$ is Bohr radius and $a_{\perp}=\sqrt{\hbar/(m\omega_{\perp})}$.

Applying the ansatz (\ref{ansatz}) in Eq.(\ref{3dgp}) we obtain the quasi one dimensional (cigar-shaped) GP equation, describing the dynamics of BEC as follows
\begin{eqnarray}
\hskip-2cm i\frac{\partial\psi(x,t)}{\partial t} &=& \left[ - \frac{1}{2}\frac{\partial^2}{\partial x^2} + \frac{1}{2} M x^{2} + \tilde{V}_{\mathrm{opt}} \right]\psi(x,t) \nonumber \\& & + g |\psi(x,t)|^2\psi(x,t) - \tilde{\mu}\psi(x,t)\label{gp},
\end{eqnarray}
where $g(t) = 2a_s(t)/a_B$, $M = \omega^{2}_{0}/\omega^{2}_{\perp}$, $\tilde{V}_{\mathrm{opt}} = V_{\mathrm{opt}}/(\hbar \omega_{\perp})$, $\tilde{\mu}=\mu/(\hbar\omega_{\perp})$. Here it is important to note that $x$ and $t$ are now actually dimensionless, i.e., $x\equiv x/a_{\perp}$ and $t\equiv\omega_{\perp}t$. From here onward, we will follow this dimensionless notation of $x$ and $t$. To begin with,  we would concentrate on the dynamics of the condensate in the absence of the harmonic confinement (i.e., when $M = 0$) and would assume that the BEC experiences only the optical potential. However, in the later part of our analysis, we will remove this assumption and explore the trapped scenario ($M\neq 0$) as well which is physically more relevant.
\begin{figure}[t]
\begin{center}
\includegraphics[scale=0.25]{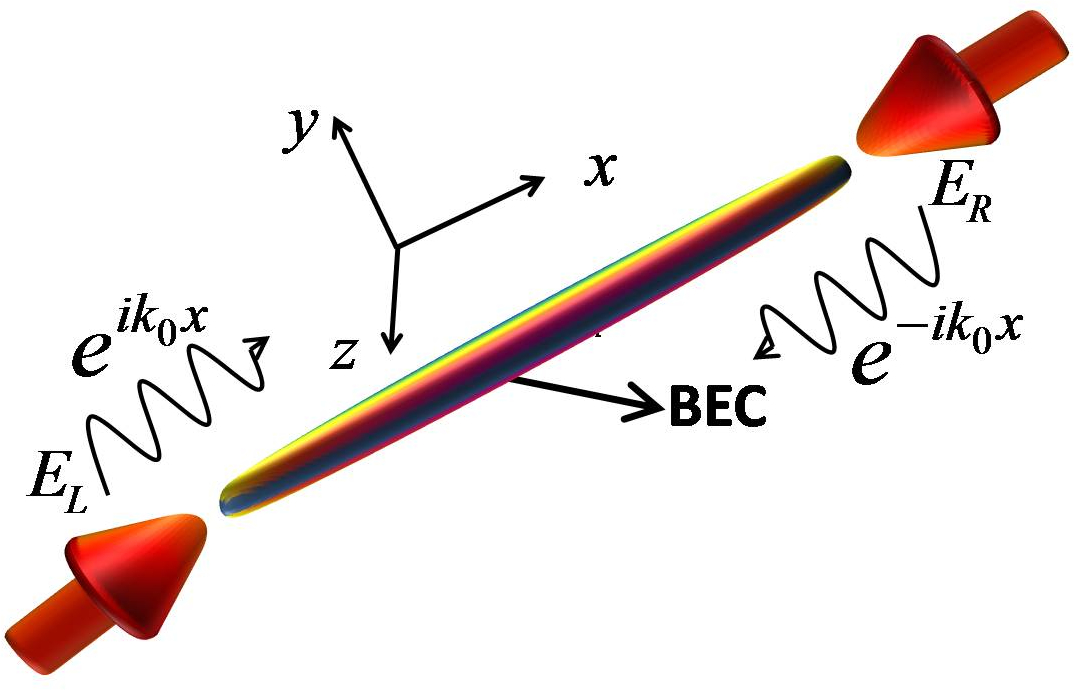}
\caption{(Color online) Schematic diagram of the physical system considered here. A cigar-shaped BEC is exposed to two counter-propagating laser beams.}
\label{schematic}
\end{center}
\end{figure}
The quasi one-dimensional BEC described above is considered to be exposed to two counter-propagating lasers of equal intensity, as illustrated in
Figure~\ref{schematic}. The lasers are far detuned from the atomic resonance such that the possibilities of atomic excitation and spontaneous emission from the atoms can be ignored. The impinging laser fields from the left ($\mathcal{E}_{L}(x,t)$) and right ($\mathcal{E}_{R}(x,t)$) can be assumed to be of the form $\mathcal{E}_{L,R}(x,t)=\left[E_{L,R}(x)e^{i\omega_lt}+\rm{c.c.}\right]\hat{e}_{L,R}$. The orthogonal polarization ensures $\hat{e}_L\cdot\hat{e}_R=0$. Further we consider that the laser field is associated with a fast varying component, i.e., $E_L(x)\equiv \tilde{E}_{L}(x)e^{ik_0x}$ propagates from the left, whereas, $E_R(x)\equiv \tilde{E}_{R}(x)e^{-ik_0x}$ propagates from the right, with $\omega_{l}$ being the frequency of the laser pulses and $k_0$ is the wave number of the incoming beams.  After adiabatically eliminating the propagation delay by considering the fact that the propagation time of the light through the atomic sample is negligible compared to any other time scale that exists in the system, the envelop of the two EM fields can be written in the form of Helmholtz equation \cite{ritsch1},
\begin{eqnarray}
\frac{d^2\tilde{E}_{L,R}(x)}{dx^2}+k_0^2 (1+\chi(x))\tilde{E}_{L,R}(x)&=&0,\label{helmoltz}
\end{eqnarray}
with $\chi(x)$ being the susceptibility of the BEC, arising as a result of nonlinear effect induced by the propagation of an intense laser beam through BEC medium and is defined as, $\chi(x)=\alpha |\psi(x)|^2/|\psi_0|^2$. The strength of atom-light coupling is $\alpha$ with $|\psi_0|^2$ refers to the background density of the condensate. The background density is defined through Thomas-Fermi (TF) approximation (by ignoring the kinetic energy contribution), which explicitly turns out as $|\psi_0|^2=\sigma_0=\tilde{\mu}/g$.

For a known spatial distribution of the incoming laser pulses, one can determine the optical potential experienced by the atoms in the condensate as,
\begin{eqnarray}
\tilde{V}_{\mathrm{opt}}(x) &=&  \frac{\alpha}{\hbar\omega_{\perp}} \left(|\tilde{E}_L(x)|^2+|\tilde{E}_R(x)|^2\right) = \tilde{\alpha} |\tilde{E}(x)|^2,
\end{eqnarray}
where we assume that the intensity of the left and right laser pulses are the same, $|\tilde{E}_L(x)|^2 = |\tilde{E}_R(x)|^2 \equiv |\tilde{E}(x)|^2$ and $\tilde{\alpha} = 2\alpha / (\hbar\omega_{\perp})$. The model described above provides new insights in the study of various nonlinear phenomena due to its extra tunability and possibility to electromagnetically carve out the sign and strength of the underlying optical potential on demand. For notational convenience, we now drop the {\it `tilde'}, wherever applicable. In the following section, we illustrate the exact analytical solution of the model system for different regimes and analyze its stability criterion.

\section{Solutions}\label{solu}

In this section, we aim to investigate the possibilities of localized mode generation in the condensate due to the light-matter interaction. Apparently, the analysis appears analogous to the earlier studies on the two-component BEC systems (\cite{kevre,das1} and references there in). However the physical system studied here is quite different as described above. Moreover, through the current formalism we observe (a) generation of different optical potentials (b) competition of atom-atom and light-atom interaction which is observed to lead to significant changes in the nature of effective interaction, (c) generation of dark soliton like mode in the atomic condensate and (d) signature of bound state formation for dark soliton.

\subsection{Localized Modes}

In order to obtain localized solutions of the modified GP equation, it is important that we approximate the pulse profile by a localized function as shown in Eq. (\ref{pulse}). This particular pulse profile is quite common in case of pico/femto second lasers \cite{joachain}. Furthermore, in order to reduce the computational difficulty, we choose to move to the center of mass (CoM) frame from the laboratory frame. In the CoM frame, the condensate wavefunction can be considered as, $\psi(x,t) = \sqrt{\sigma_{a}(\eta)} e^{i\chi_a(\eta)}$ with $\eta = x - u t$; $u$ being the condensate velocity in CoM frame, $\sigma_{a}$ is the density and the condensate and $\chi_{a}$ refers to the nontrivial phase of the BECs. Readers should note that this nontrivial phase actually mimics a velocity potential, which in turn is directly related to the irrotational velocity of the condensate, $v = \partial \chi_{a}/\partial \eta$. In a similar manner, the laser pulse profile can also be transformed as, $E(x) = \sqrt{I(\eta)}$, with,
\begin{eqnarray}
I(\eta) & = & \sigma_{0}\cos^{2}\theta\, \textrm{sech}^{2}\left(\frac{\cos\theta}{\zeta}\eta\right)\label{pulse}
\end{eqnarray}
is the intensity of the pulse profile. Here, $\sigma_0$ is the TF  density or the background density of the condensate, $\theta$ is the Mach angle, and $\zeta$ is the healing length. The profile is expressed in the same frame of reference as that of the BEC center of mass. The transformation to the CoM frame allows us to treat the dynamics of the pulse profile as well as the BEC in the same footing. In what follows, we would show graphically that the laser profile actually create an optical potential of bell type or P\"{o}schl-Teller type. Applying this ansatz in Eq.(\ref{gp}) we solve the coupled equations, Eq. (\ref{gp}) and (\ref{helmoltz}) simultaneously. The imaginary part of the GP equation mimics the current conservation, which can be written in the following compact form: $v = u (1-\sigma_0/\sigma_a)$.

\begin{figure}\centering
\includegraphics[scale=0.22]{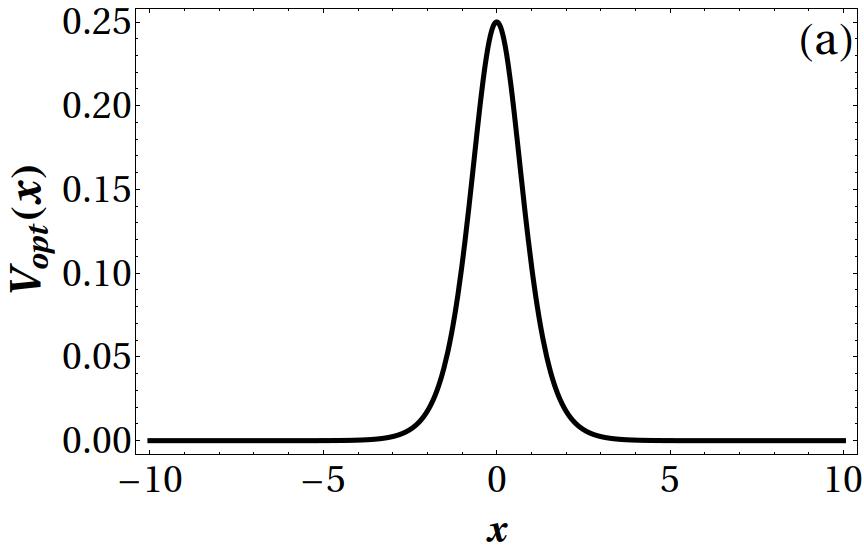}
\includegraphics[scale=0.22]{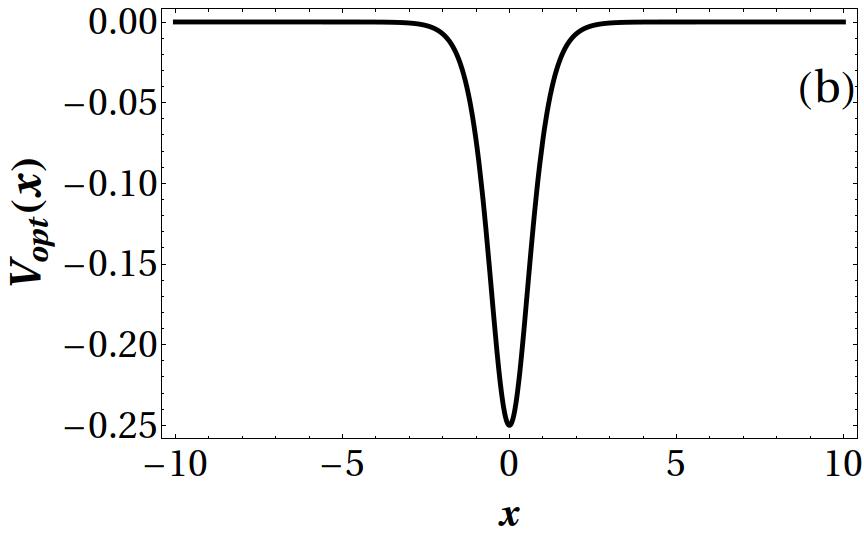}
\caption{(a) The effective bell type optical potential generated due to the light-matter interaction. Here, both $g$ and $\alpha$ are repulsive (positive), but $\alpha<g$. We have used an arbitrary value for light-matter coupling strength i.e. $\alpha=0.5$. (b)  The effective P\"{o}schl-Teller type optical potential is generated due to the light matter interaction. Here $g$ ($\alpha$) corresponds to a  repulsive (attractive) interaction. Here we consider $\alpha=-0.5$. In both the figures, $x$ is in the units of $a_{\perp}$, $V_{\textrm{opt}}(x)$ in the units of $\hbar\omega_{\perp}$ and $g=1$.}\label{barrier}
\end{figure}

The real part of the GP equation yields,
\begin{eqnarray}
-\frac{1}{4}\sigma_a\frac{d^2\sigma_a}{d\eta^2}+\frac{1}{8}\left(\frac{d\sigma_a}{d\eta}\right)^2+g\sigma_a^3 &+& \alpha I \sigma_a^2 -\left(\frac{u^2}{2}+\mu\right)\sigma_a^2\nonumber \\ &+& \frac{u^2}{2}\sigma_0^2=0.\label{real}
\end{eqnarray}
The solution of the above equation is obtained as,
\begin{eqnarray}\sigma_a &=& \sigma_0\left(1-\cos^2\theta\cosh^{-2}\left(\frac{\cos\theta}{\zeta}\eta\right)\right).\label{sol}
\end{eqnarray}
The healing length is obtained as $\zeta=1/\sqrt{(g-\alpha)\sigma_0}$ and the wave number $k_0^2=\kappa\sigma_0\left(\cos^2\theta-2\right)/2$, with $\kappa=2\left(-g+\alpha\right)$. The Mach angle would be defined as the ratio of soliton velocity and the sound velocity such that $\sin\theta=u/u_s$, where the sound velocity $u_s=\sqrt{(g-\alpha)\sigma_0}$. As the healing length is always positive, we explore several possible interaction regimes corresponding to  the allowed relations between atom-atom coupling strength $g$ and light-matter coupling strength $\alpha$, which are required to satisfy $|g-\alpha|>0$. Let us now elaborate each of these possibilities:
\begin{itemize}
\item{Case I}:  $g>0$, $\alpha>0$, but $\alpha<g$; this implies that atom-atom interaction as well as light-matter interaction is repulsive. In this case, the effective optical potential behaves like a barrier as illustrated in Figure~\ref{barrier}a.
\end{itemize}
\begin{itemize}
\item{Case II}:  Contrary to the previous case, for $g>0$ and $\alpha<0$ the effective optical potential turns out to be P\"{o}schl-Teller type which can support a bound state formation (see Figure~\ref{barrier}b). In both the cases for $\alpha\rightarrow0$ the coherence length and the sound velocity takes the usual form for homogeneous BEC.
\end{itemize}

\begin{itemize}
\item{Case III}: For the sake of completeness, we can mention situations like an attractive $g$ (light-matter interaction both attractive and repulsive) however, it directly leads to negative background density. Therefore, we limit our analysis for the above two cases only.
\end{itemize}
\begin{figure}\centering
\includegraphics[scale=0.21]{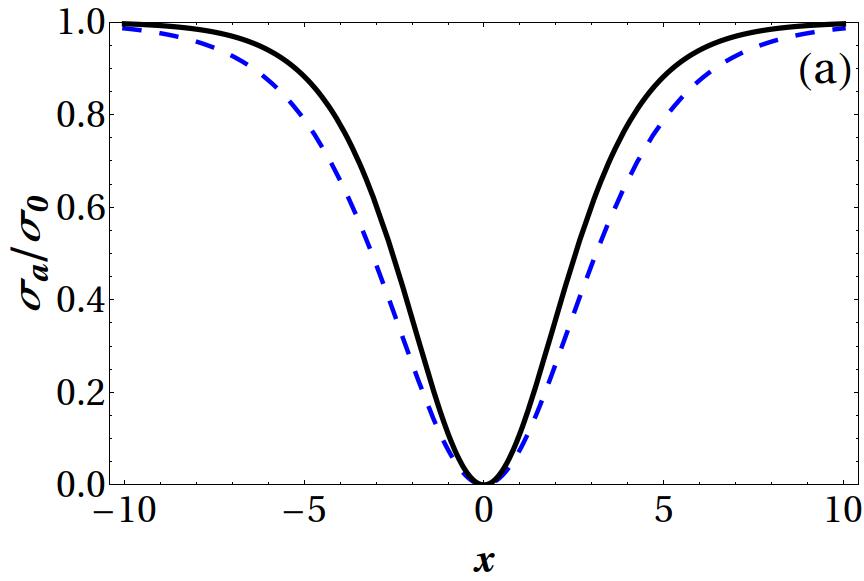}
\includegraphics[scale=0.21]{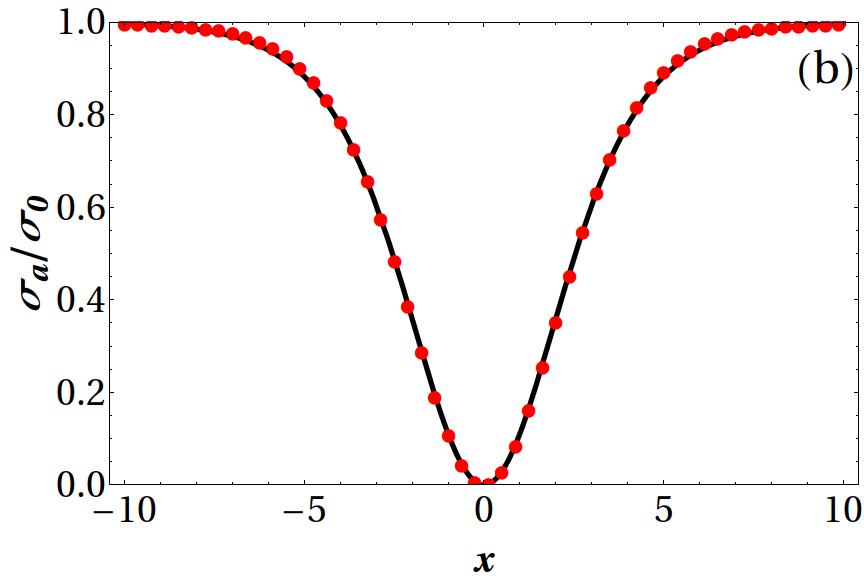}
\caption{(Color online) (a) Variation of the dark soliton width due to attractive and repulsive photon-atom coupling. The dashed (blue) and solid (back) line correspond to $\alpha=0.2g$ and $\alpha=-0.2g$, respectively. We used an arbitrary repulsive value for $g$ which reads $g=1.25$. Here, $x$ is represented in the units of $a_{\perp}$. (b) Comparison of numerical (red, dotted) and analytical (black, solid) results for repulsive atom-photon interaction $\alpha=-0.2g$.}
\label{density}
\end{figure}

To visualize the different regimes of the light-matter interaction involved in the physical system of our interest, Eq. (\ref{sol}) is plotted in Figure~\ref{density}a for two different regimes. In the figure, the dashed blue and solid black line correspond to $\alpha=0.2g$ and $-0.2g$ respectively. Here, we note that the total number of atoms forming the solitons is less than the number of atoms in the condensate, due to the presence of the Thomas-Fermi background $\sigma_{0}$ \cite{das1,jackson1998solitary,jackson,komineas2003nonlinear,komineas2003solitons,jackson2006bose}.

We corroborate our analytical result with numerical simulation in Figure~\ref{density}b where the solid line depicts the analytical solution for $\alpha=-0.2g$ and the numerically obtained solution is represented via dotted line. The numerical simulation was carried out by using  split-step Crank-Nicholson (CN) method by discretizing in space and time \cite{muruganandam2009fortran}. Due to the localized nature of the wavefunction obtained analytically, we consider $\tanh$-profile as an initial wavefunction, whereas, the boundary conditions are evolved with the initial wavefunction. While discretising, we choose $dz = 0.025$ and $dt = 0.00002$, satisfying the CN criterion: $dt/dz^2  < 1$. We evolve the initial wavefunction for $200000$ iterations and obtained an intermediate solution. This intermediate solution is further evolved for $20000$ iterations, in order to check the convergence of the wavefunction and thus obtained the final solutions. Figure~\ref{density} (b) clearly indicate the appreciable agreement of the analytical and numerical results.

Nevertheless, the figures do not carry the necessary information regarding the stability of the solitons and the energy transfer between the BEC and laser pulses. Therefore, it is extremely important to know the stability criterion of the obtained solution and 
associated energy exchange details between the BEC and the laser pulses. Hence, we explicate these issues here onward.

\subsection{Stability analysis}
In the previous section, we have elaborated on the existence of localized modes for a wide range of parameters. The main objective of this section is to investigate the stability of the obtained solitonic modes.
Physically, our condensate yields a dark soliton due to the bright soliton-like pulse profile via light matter interaction. The well-known Vakhitov-Kolokolov (VK) criterion \cite{vakhitov1973stationary} ensures the stability of solitons for the 1D GP equation for bright solitons. However, here we use a method that can be viewed as a modified version of the soliton perturbation theory (see \cite{kivshar1989} and references therein) considering a slow evolution of solitons near the instability threshold. An extended analysis of the method can be found
in Ref.~\cite{barashenkov}, where, it is proven that the negative gradient of generalized momentum with respect to the soliton velocity is a sufficient stability criterion for dark solitons. One can also find application of similar method in Refs.~\cite{hadi,pendse}. This is on contrary to the bright solitons where the nature of particle number versus chemical potential graph defines the stability \cite{vakhitov1973stationary}. In what follows, we have followed the method described in ~\cite{barashenkov} and have referred to the corresponding stability criterion as VK-like criterion.
To investigate the stability of the dark soliton solution, we calculate the renormalized momentum,
\begin{eqnarray}
P_a &=& -i\int\psi^*\frac{\partial\psi}{\partial x}dx =\sigma_0\zeta u_s\left(\pi\frac{u}{|u|}-2\theta-\sin2\theta\right).\label{momentum}
\end{eqnarray}
The stability criterion tells us that the soliton will be stable if $\mathcal{P}=\partial P_a/\partial u<0$ \cite{barashenkov}. In the current scenario the stability condition yileds, $\mathcal{P}=-4\sigma_0\zeta\cos\theta$ which clearly suggests that the dark solitons are stable
as long as $(g-\alpha)>0$. This condition clearly points to the second scenario (Case-II) mentioned in the earlier passages as the region of stability.
\begin{figure}[t]
\centering
\includegraphics[scale=0.27]{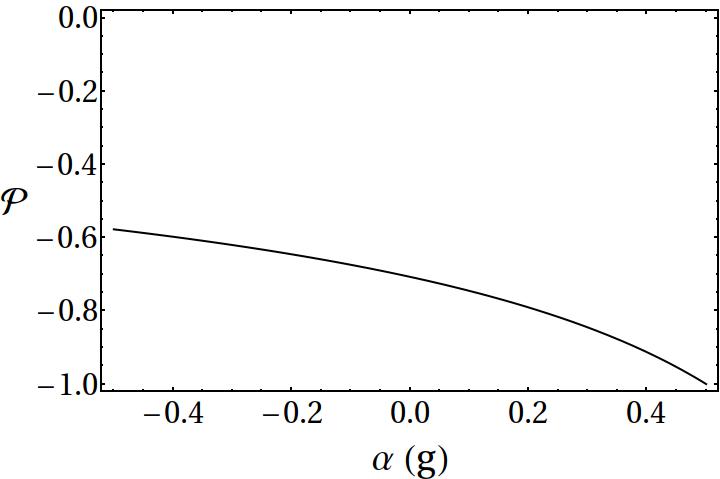}
\caption{Variation of $\mathcal{P}$ with $\alpha$ while we swipe $\alpha$ from attractive to repulsive regime. Here, $g = 0.5$ is repulsive and $|\alpha|<g$, $\theta=\pi/4$. Here, $\mathcal{P}$ and $\alpha$ are scaled by $4\sqrt{\sigma_{0}}$ and $g$, respectively. }
\label{stability}
\end{figure}
Figure~\ref{stability} clearly shows the stable regions for the obtained localized solution. In the figure, the instability parameter $\mathcal{P}$  is shown as a function of the atom-light interaction parameter $\alpha$, when $g>0$. For all values of $\alpha$, subjected to the condition $g>\alpha$, we observe that the instability parameter is always negative, indicating that the obtained solution is stable within this parameter regime.

It is also worth noting that the competition between $g$ and $\alpha$ can impart significant effect on the pulse profile and we realize that $(g-\alpha)>0$ condition again plays important role if we look forward for stable pulse profile.

\section{Energy of the Solitonic Mode}\label{ene}

To analyze the solitonic mode further, which is gray in nature, under the influence of the counter-propagating laser pulses, we minimize the energy and compute the energy functional as,
\begin{eqnarray}
E_{a}&=&\frac{1}{2}\int\left(\frac{\partial\psi^*}{\partial x}\right)\left(\frac{\partial\psi}{\partial x}\right)dx+\frac{g}{2}\int\left(\psi^*\psi\right)^2dx\nonumber\\&& + \alpha\int\left(E_L^*E_L\right)\left(\psi^*\psi\right)dx-\mu\int\left(\psi^*\psi\right)dx.\label{e_functional}
\end{eqnarray}
Subsequently, using Eqs. (\ref{sol}) and (\ref{e_functional}) and subtracting the background we obtain
\begin{eqnarray}
E_{a}&=&\frac{4}{3}g\sigma_0^2\zeta\cos^3\theta+2\alpha\sigma_0^2\zeta\left(\cos\theta-\frac{2}{3}\cos^3\theta\right).\label{energy}
\end{eqnarray}
Corresponding momentum has already been obtained in Eq.(~\ref{momentum}).
\begin{figure}[h]
\centering
\includegraphics[scale=0.25]{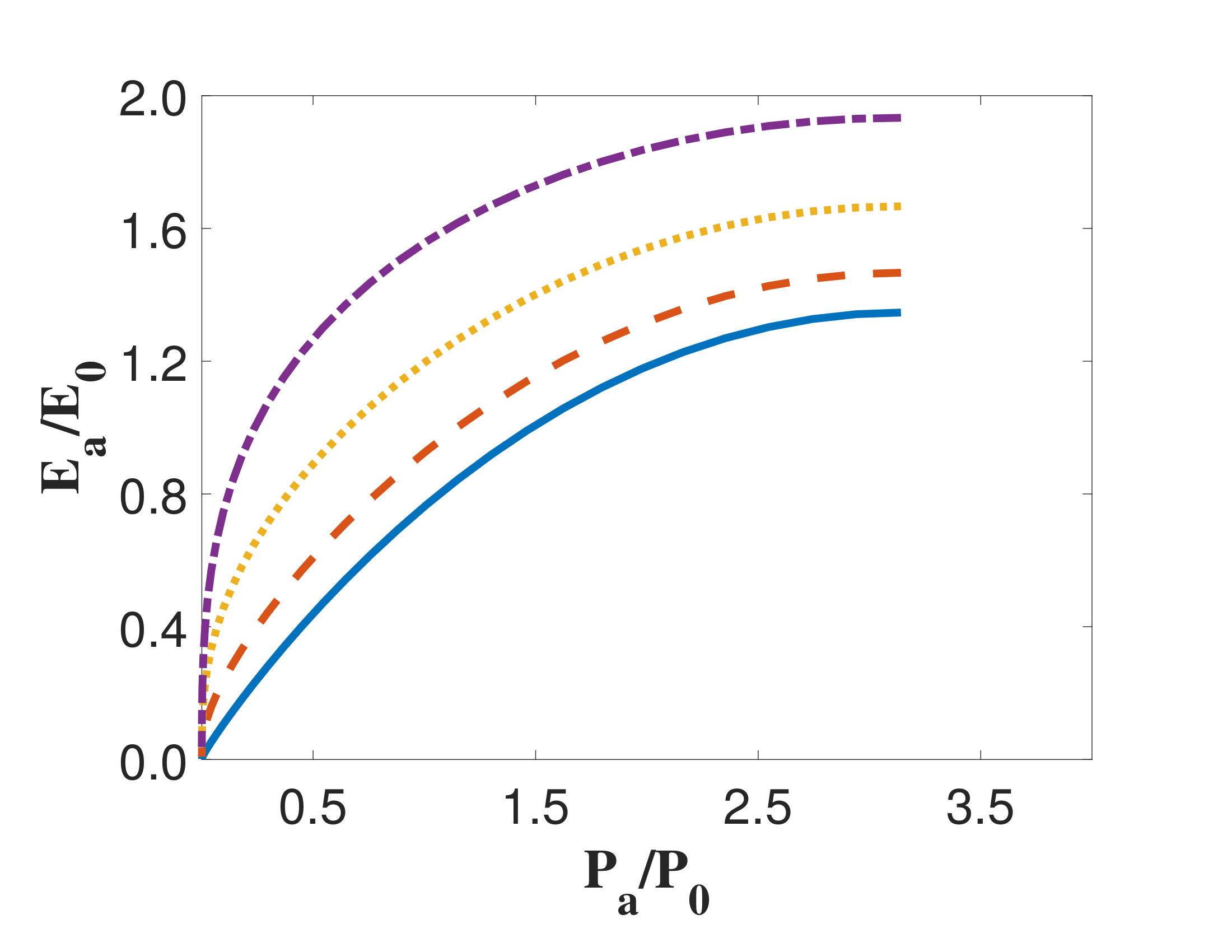}
\includegraphics[scale=0.25]{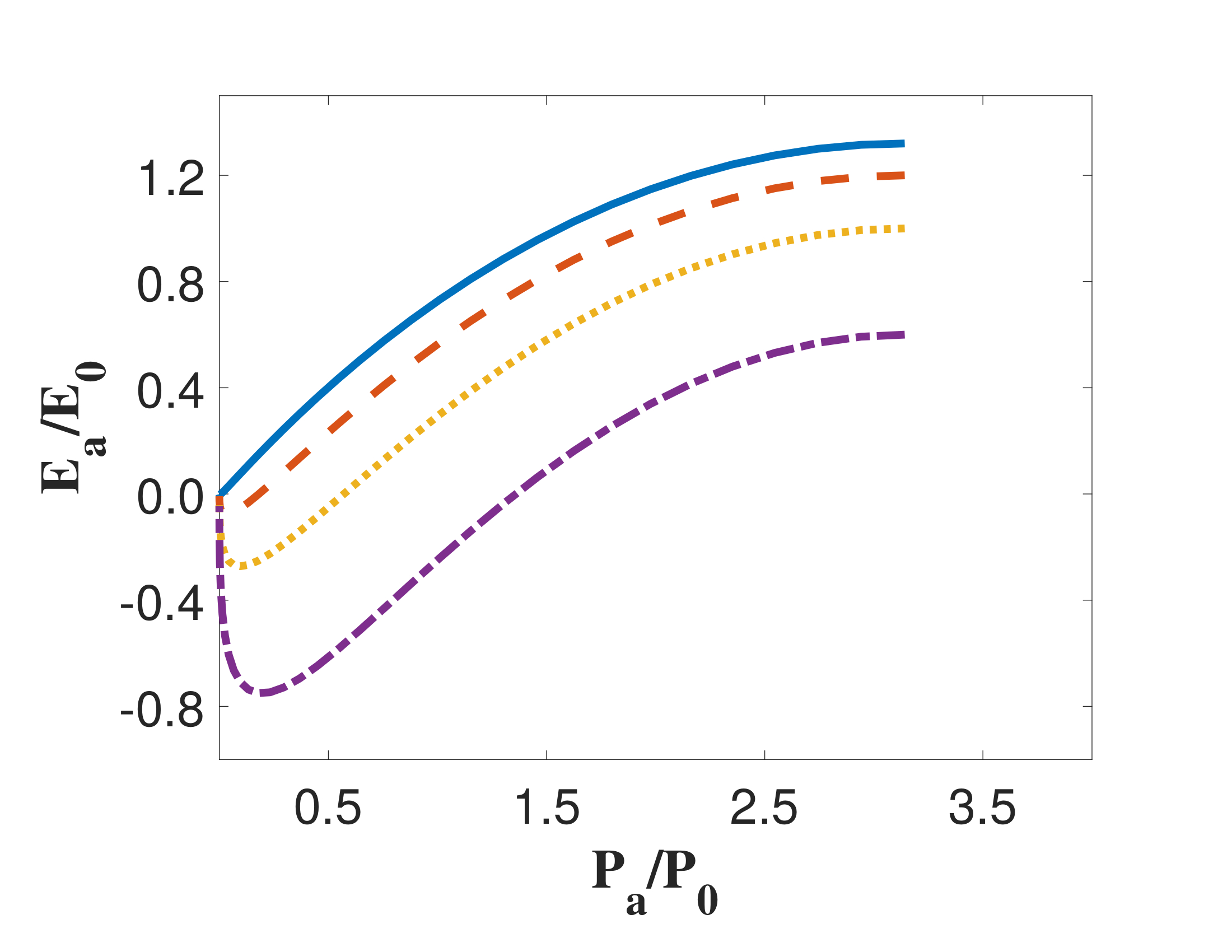}
\caption{(Color online)  The energy momentum dispersion of the matter wave when the repulsive BEC is subjected to an effective optical potential of bell type (Case I),   P\"{o}schl-Teller type (Case II). Here $E_0=\sigma_0^2\zeta$ and $P_0=\sigma_0\zeta u_s$. In (a) solid line corresponds to $\alpha=0.02g$, dashed line is for $\alpha=0.2g$, the dotted line depicts $\alpha=0.5g$ and dispersion with $\alpha=0.9g$ is presented using dashed dotted line. Similarly, in (b)  solid line corresponds to $\alpha=0.02g$, dashed line is for $\alpha=0.2g$, the dotted line depicts $\alpha=0.5g$ and dispersion with $\alpha=1.1g$ is presented using dashed dotted line.}
\label{dispersion}
\end{figure}

\begin{figure*}[t]
\begin{center}
\includegraphics[scale=0.2]{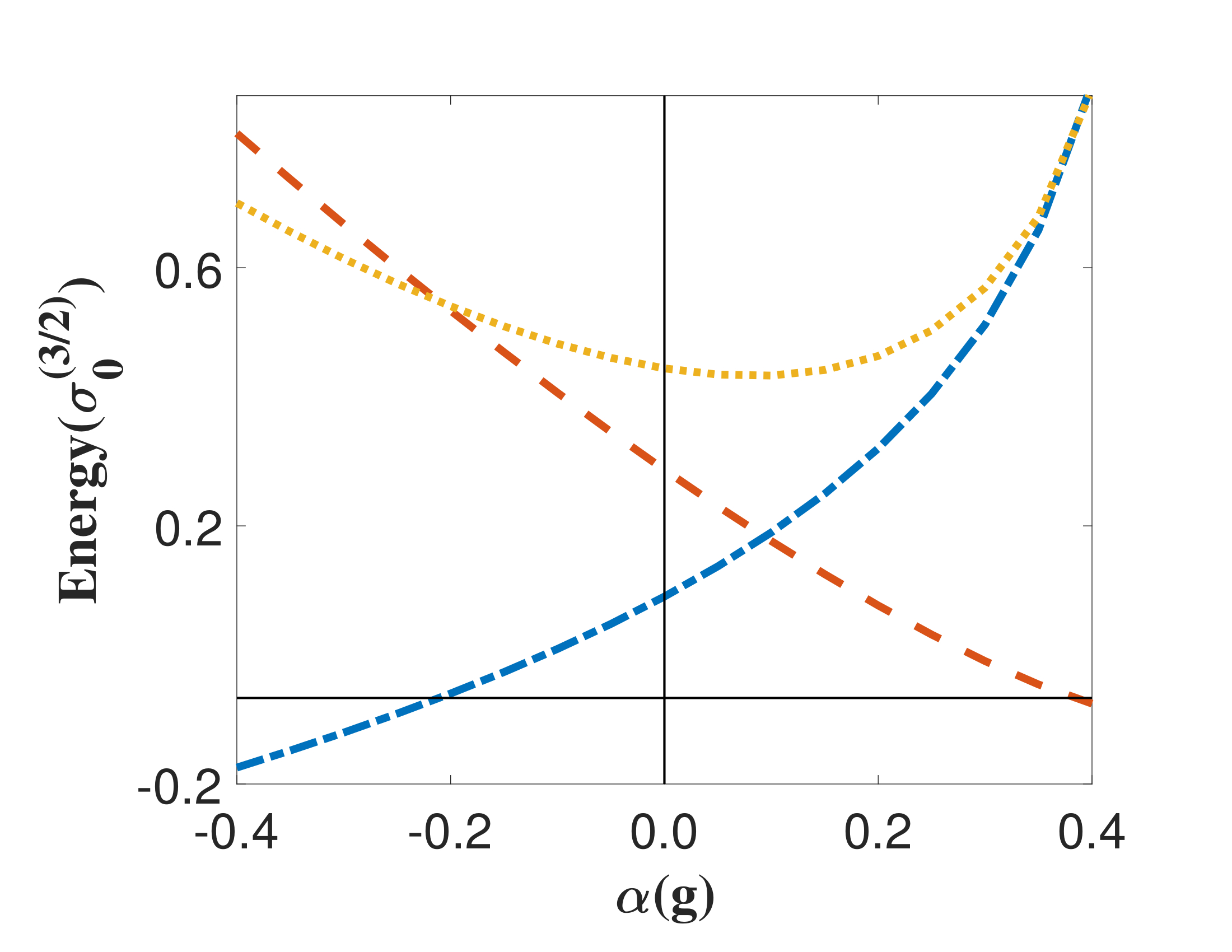}
\includegraphics[scale=0.2]{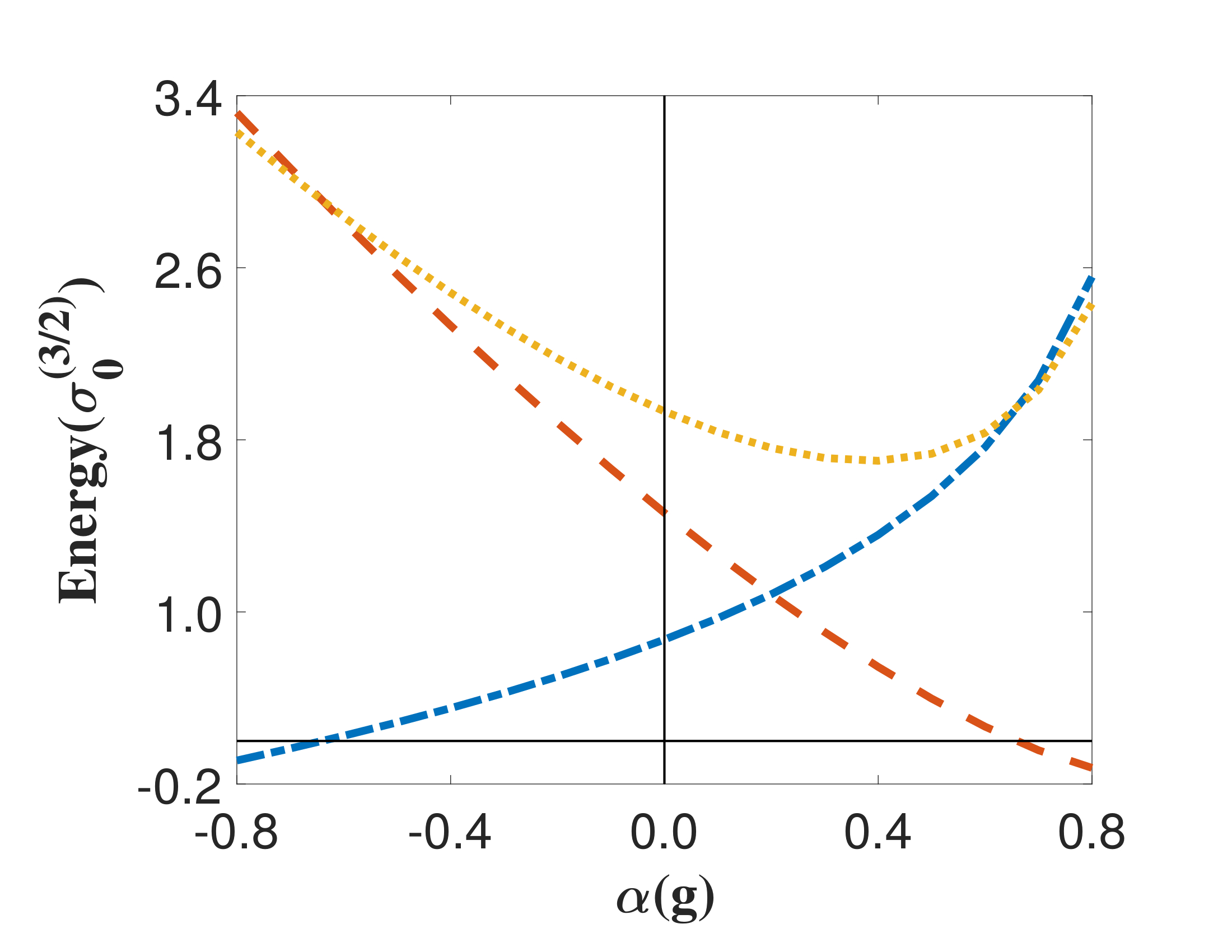}
\includegraphics[scale=0.2]{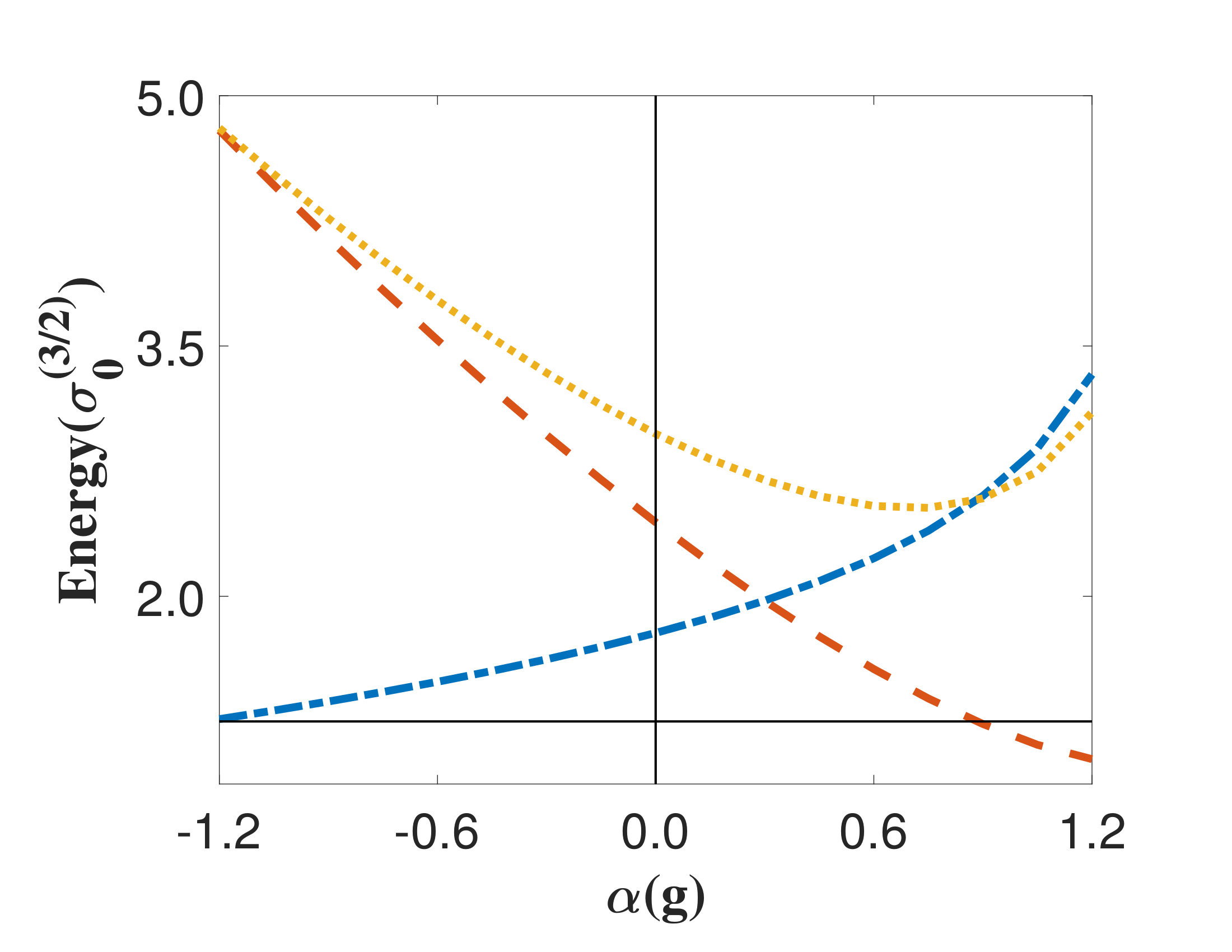}\\
\hfill (a)\hfill\hfill (b)\hfill\hfill (c)\hfill\hfill
\caption{(Color online) Variation of energy of the solitonic mode and photonic mode as light-matter interaction is tuned ($\alpha$). In all the figures the dashed-dotted line represents soliton energy, dashed line is for pulse energy and the sum of these two energies is illustrated through dotted line. Across the figures $\alpha$ is varied as a function of $g$, to be precise from $-0.8g$ to $0.8g$. The value of $g$ from left to write is as follows: $0.5$, $1.0$, $1.5$. All the energies are in units of $\sigma_0^2\beta$. The plots are prepared for $\theta=\pi/4$.}
\label{eatom}
\end{center}
\end{figure*}

The energy vs momentum dispersion curves for the above mentioned cases are depicted in Figure~\ref{dispersion}. In Figure~\ref{dispersion}a, the dispersion curve is given for Case I and we observe the usual Lieb dispersion \cite{lieb,jackson}. Nevertheless, the maximum energy corresponding to $\pi$ momentum differs considerably based on the strength of $\alpha$. The situation is quite different for repulsive BEC with attractive light matter interaction (see Figure~\ref{dispersion}b). The competition between $g$ and $\alpha$ leaves signature of solitonic bound state formation at low momentum with sufficiently high $\alpha$ (when $|\alpha|$ is of the order of $\sim0.2g$ or above). However, at high momenta the system tries to follow its usual path.

The formation of the bound state observed in Figure~\ref{dispersion}b in case of attractive atom-photon interaction, can be well explained from the low energy scattering theory \cite{sakurai1995modern}. It is evident from Figure~\ref{dispersion}b that bound state occurs in the small momentum regime, when the light-matter interaction strength ($\alpha$) is attractive. In case of repulsive atom-atom interaction (positive $\alpha$), we observed the usual Lieb dispersion profile. However, when we move to the negative value of $\alpha$ (attractive light-matter interaction), we observed the formation of bound state. This results in a transition of the system from the weak interaction regime to strong interaction regime, which favors the bound state formation \cite{chen2014pseudogap}. It is worth mentioning that in ultra-cold atomic systems, such bound states can easily be controlled in a coherent manner and may be useful for information storage and retrieval. Furthermore, the phase of the condensate depends on the density, whereas, the phase of the laser pulse profile is of kinematic origin. This relative phase difference between the condensate and that of laser pulse is important and has useful application in coding information, where the information can be stored in the form of bound states \cite{agarwal2003slow}.

It is worth noticing that in Case II, we do not have any constraints on the strength of $g$ and $\alpha$ except $g>0$. Further, the dispersion diagram for this regime indicates bound state formation for solitons. This motivates us to probe this interaction regime further. To reveal the interplay between the two interactions, where one ($g$) tries to delocalize the solitons and the other ($\alpha$) tries to localize them. It is important that we study the energy transfer mechanism between light and matter. For this purpose, we calculate the pulse energy ($E_P$) which is defined as 
\begin{eqnarray}
  E_P &=& \frac{k_0^2}{2}\int(1+\chi(x)) E_L^*E_L dx\nonumber\\
  &=&k_0^2\sigma_0\zeta\cos\theta + \frac{2}{3}\alpha k_0^2\sigma_0\zeta\cos\theta\left(2\cos^2\theta-3\right),
\end{eqnarray}
and plot
$E_a$  and $E_P$ with varying $\alpha$ (see Figure~\ref{eatom}) for three different values of atom-atom interaction strength. In general, we observe that pulse energy decreases as an effect of tuning $\alpha$ from attractive to repulsive, whereas soliton energy is found to increase for the same situation. This clearly suggests the occurrence of
energy transfer from the laser pulse to soliton.

In Figure~\ref{eatom} the energy is normalized by $\sigma_0^2\beta$ where $\zeta=\beta/\sqrt{(g-\alpha)}$ and $\beta=1/\sqrt{\sigma_0}$.
In the left most figure (Figure~\ref{eatom}a), we again observe the signature of solitonic bound state formation for attractive $\alpha$ while $g$ is moderately weak ($g=0.5$) (as pointed out in Figure~\ref{dispersion}b as well). However, as we increase the repulsive strength of $g$ this signature quite naturally disappears (see Figure\ref{eatom}b and Figure\ref{eatom}c). The threshold value for $\alpha$ ($\alpha_c$), for which soliton energy experiences the level crossing by becoming positive from negative, can be noted as,
\begin{eqnarray}
&&\alpha_c=-\frac{1}{2 (-2 + \cos(2 \theta))^2} \Big(4 g \cos^{4}\theta + \nonumber\\
&& \left(2 g \cos^{4}\theta (36 + 3 g + 4 (-8 + g) \cos(2 \theta) + (4 + g) \cos(4 \theta))\right)^{1/2}\Big).\nonumber\\
\label{alpha}
\end{eqnarray}
Using Eq.(\ref{alpha}) we see that $\alpha_c=-0.2g, -0.6g$ and $-1.2g$ for $g=0.5, 1.0$ and $1.5$, respectively which corroborates well with the zero crossing of solitonic energy in Figure~\ref{eatom}.
Further, we observe a non-monotonic behavior of the total energy as depicted in the Fig.(6). Looking into the non-monotonic nature of the total energy more closely we realize that the in case of attractive atom-photon coupling, the energy due to the fast varying component of the laser pulse ($k_{0}$) dominates and hence, the total energy initially decreases. However, in the repulsive regime, the contribution from the atom-atom interaction and atom-photon coupling dominates and hence we observe a sharp increase in energy. The fact that the attractive $\alpha$ supports the bound state formation of solitons, indicating the possibility of the presence of a self-bound quantum droplet state, which have been studied in a number of papers in recent times \cite{petrov2015quantum,petrov2016ultra,barbut,debnath2020investigation}. This also opens up scope for further investigation on the formation of the droplets through the light-matter interaction.

\section{Conclusion}\label{con}
In summary, we have investigated the effect of two orthogonally polarized, counter propagating laser beams on atomic condensate. The pulse profile has a localized structure and shining them on the condensate results in generation of localized profile in the form of solitons. We have observed that the interplay between atom-atom coupling strength $g$ and light-matter coupling strength $\alpha$, leads to different regime where these solitons are found to exist. These localized structures are found to be stable only for repulsive atom-atom interaction and attractive light-matter interaction. Furthermore, we have studied the effect of light-matter interaction on Lieb mode, where we have observed the signature of solitonic bound state formation. The obtained energy diagram also indicates a transfer of energy from light pulse to atom as the light-matter interaction parameter, $\alpha$ changes its sign. In general, the light energy is found to gradually transferred to the atomic condensate as we move from negative to positive $\alpha$. Since the attractive $\alpha$ connects with a  P\"{o}schl-Teller type optical potential, it supports solitonic bound state formation.

We would like to emphasize that our analysis has revealed that the light-BEC interaction can be investigated in multiple regimes and meticulous understanding of physical phenomena happening at each of these regimes requires urgent attention.
We believe that our current study would provide some insight into this issue and initiate an intense effort to examine novel aspects of light-matter physics.

\section*{Acknowledgments}
Authors acknowledge insightful discussions with Prasanta K. Panigrahi. PD also acknowledges Indian Institute of Technology Delhi for providing the facilities, where this work has been started. AK thanks Science and Engineering Research Board (SERB), Department of Science and Technology (DST), India for the support provided through the project number CRG/2019/000108. AP thanks Department of Science and Technology (DST), India for the support provided through the project number EMR/2015/000393.

\section*{Authors contributions}
AP conceived the presented idea. PD and AK carried out the theoretical formalism, performed the analytical calculations and numerical simulation. All the authors have equally contributed to the analysis of the results and the preparation of the manuscript.

\bibliographystyle{epj}
\bibliography{ms}

\end{document}